\documentclass[preprint]{aastex63}

\graphicspath{{./}{figures/}}

\usepackage{xcolor}
\usepackage[colorinlistoftodos]{todonotes}

\let\a=\alpha \let\b=\beta \let\g=\gamma \let\d=\delta \let\e=\epsilon
\let\h=\eta  \let\k=\kappa \let\l=\lambda 
\let\s=\sigma \let\t=\tau \let\u=\upsilon
\let\r=\rho
 
\let\w=\omega	   \let\G=\Gamma \let\D=\Delta  \let\L=\Lambda
\let\S=\Sigma \let\W=\Omega
\let \x = \chi \let \z = \xi

\def\be{\begin{eqnarray}}
\def\ee{\end{eqnarray}}
\def\ba{\begin{array}}
\def\ea{\end{array}}

\def \kms {\mathrm{km \, s^{-1}}}

\def \M  {\mathscr{M}}
\def \K {\mathrm{K}}

\def \earth {\oplus}
\def \S {Section }

\shorttitle{submm variability of Sgr A* in June 2019}
\shortauthors{Lena Murchikova \& Gunther Witzel}
\begin{document}

\title{Second Scale Submillimeter Variability of Sagittarius A*\\during flaring activity of 2019:\\ On the Origin of Bright Near Infrared Flares
}

\correspondingauthor{Lena (Elena M.) Murchikova}
\email{lena@ias.edu}

\author[0000-0001-8986-5403]{Lena Murchikova}
\affiliation{Institute for Advanced Study, 1 Einstein Drive,
Princeton, NJ 08540, USA}

\author[0000-0003-2618-797X]{Gunther Witzel}
\affiliation{Max-Planck-Institut f\"ur Radioastronomie, Auf dem H\"ugel 69,
53121, Bonn, Germany}

\begin{abstract}

In 2019, Sgr A*  -- the supermassive black hole in the Galactic Center -- underwent unprecedented flaring activity in the near infrared (NIR), brightening by up to a factor of 100 compared to quiescent values. Here we report ALMA observations of Sgr A*'s continuum variability at 1.3 mm (230 GHz) -- a tracer of the accretion rate -- conducted one month after the brightest detected NIR flare and in the middle of the flaring activity of 2019.
We develop an innovative light curve extraction technique which (together with ALMA's excellent sensitivity) allows us to obtain the light curves which are simultaneously of high time resolution (2 seconds) and high signal-to-noise ratio ($\sim 500$). We construct an accurate intrinsic structure function of the Sgr A* submm variability, improving on previous studies by about two orders of magnitude in timescale and one order of magnitude in sensitivity.
We compare the June 2019 variability behavior with that of 2001-2017, and suggest that the most likely cause of the bright NIR flares is magnetic reconnection.
%The 230 GHz flux behave in agreement with the mechanism producing the flaring activity of 2019 proposed by \cite{Murchikova2021}.

\end{abstract}



\section{Introduction} \label{sec:intro}

Sagittarius A* (Sgr A*) is a non-thermal and variable source associated with a black hole at the center of our Galaxy. It has been regularly monitored for over 25 years and shows variability and flaring activity on the scales between minutes to hours across the electromagnetic spectrum -- radio, millimeter (mm), sub-millimeter (submm), near infrared (NIR), and X-ray \citep{Witzel2020,Bower2015,Dexter2014}. The NIR activity is particularly well studied due to regular monitoring of stellar orbits around the Galactic Center \citep{Genzel10, Morris12, Witzel2018, Gravity2020}. 

On May 13, 2019, \citet{2019ApJ...882L..27D} detected
an unprecedentedly bright NIR flare, during which the flux reached at least 100 times the typical Sgr A*'s quiescent value (twice as bright as the strongest flare previously recorded). Sgr A* continued producing bright flares until (at least) the end of the year. 
\citet{2019ApJ...882L..27D} and \citet{Gravity2020} suggested that the flares might be connected to the recent passage of the S0-2(S2) star. \cite{Murchikova2021} argued against the S0-2 connection and showed that the time of the flaring activity is consistent with the nearly simultaneous  arrival of material shed by G1 and G2 objects (formerly known as clouds) near their pericenter passage. In either scenario, the increased mass accretion rate stimulates production of bright NIR flares.

The sub-millimeter flux density of Sgr A* is generally believed to be a tracer of the accretion rate, as it is dominated by a population of thermal electrons (e.g. \citealt{Yuan2014,Bower2019}).
If the flaring activity of 2019 correlates with an increased accretion rate, it should have been accompanied by a higher submm flux. Here we report 500 minutes of observations of Sgr A*'s continuum variability at 230 GHz (1.3 mm), spread over five epochs in June 2019, which were conducted by the Atacama Large Millimeter/submillimeter Array (ALMA). 
%-- one month after the brightest ever detected near-infrared flare and in the middle of the flaring activity of 2019 -- conducted by the 

For this work we develop a light curve extraction algorithm based on the model fitting in the $(u,v)$-plane. It allows us to achieve 2-second time resolution with a signal-to-noise ratio of about 500 (surpassing the existing literature by at least an order of magnitude). 

The paper is organized as follows. In Section \ref{sec:obs} we present the details of the observations and the light curve extraction algorithm. In section \ref{sec:results} we discuss the properties of the light curves we obtained. In Section \ref{sec:discussion} we compare our results with the variability studies in the literature. We present conclusions in Section \ref{sec:conclusion}. %For conversion from arcsec to au we use the distance to Sgr A* as $8$ kpc. The Schwarzschild radius of a $4 \times 10^6 \Msun$ black hole, like Sgr A*, is $0.1$ au.

\section{Observations and data analysis}\label{sec:obs}

Our data were obtained in ALMA Cycle 6 with 51 antennas in the C43-9/10 configuration for project 2018.1.01124.S (PI Murchikova). Seven observations were conducted between June 12 and June 21 2019 (Table \ref{tab:data}).
Observations span frequencies between 228.55 GHz and 233.8 GHz, with four 1.875 GHz spectral windows centered on 229.50 GHz, 230.95 GHz, 231.90 GHz, and 232.85~GHz. The central frequency is 231.2 GHz.
The achieved resolution is 0.025 arcsec. The achieved sensitivity is 0.04 mJy over a 200 $\kms$ frequency range. 
The bandpass calibrator, phase calibrator and check source are J1924--2914, J1744--3116, and J1752--2956, respectively. 
For the data reduction, data processing, and light curve subtraction we used the Common Astronomy Software Applications package (CASA), version 5.7. For calibration and data reduction we used the script provided by the North American ALMA Science Center at the National Radio Astronomy Observatory.

We phase self-calibrated each observation independently using line-free channels. 
As a model of the source for each iteration of self-calibration we fitted a narrow Gaussian to the location of the black hole in the $(u,v)$-plane using the CASA task \verb|uvmodelfit|.

\begin{deluxetable}{ccccccc}
\tabletypesize{}
\tablecaption{The June 2019 observations} \label{tab:targets}  
\tablewidth{0pt}
\tablehead{
\colhead{obs} & \colhead{date} & \colhead{start time} & \colhead{end time} & \colhead{mean} &  \colhead{$\sigma_{obs}$} %\colhead{flux} &
\\
\colhead{id} & \colhead{UTC} & \colhead{UTC} & \colhead{UTC} & \colhead{Jy} &  \colhead{mJy} %\colhead{var}  &
}
\startdata
0 & 12-Jun-2019 & 03:36:36.5 & 04:51:03.9 & 4.62 & 9.5 \\ % 0.23 & & 25, 27, 29, 31 
1 & 13-Jun-2019 & 07:11:55.6 & 08:25:06.7 & 3.14 & 6.9 \\ % 0.18 & & 57, 59, 61, 63
2 & 14-Jun-2019 & 06:25:00.6 & 07:38:28.5 & 3.33 & 15.5\\ % 0.43 & & 89, 91, 93, 95
3 & 20-Jun-2019 & 07:15:22.9 & 08:28:25.0 & 3.58 & 2.8 \\ % 0.24 & & 121, 123, 125, 127
4 & 21-Jun-2019 & 02:35:05.7 & 03:49:47.6 & 4.03 & 4.0 \\ % 0.16 & & 153, 155, 157, 159
5 & 21-Jun-2019 & 04:17:04.5 & 05:31:14.7 & 3.70 & 4.6 \\ % 0.55 & & 185, 187, 189, 191
6 & 21-Jun-2019 & 05:58:35.3 & 07:12:23.3 & 3.81 & 4.3 % 0.40 & & 217, 219, 221, 223
\enddata
\tablecomments{The seven ALMA observations of the Sgr A* conducted in June 2019. Observation number, epoch, average continuum flux, 
and observation uncertainty derived from data
%and it is in agreement with thermal rms noise 
are given for each observation.}
\label{tab:data}
\end{deluxetable}

\begin{figure*}
\vspace{-0.0cm}
\centering
\begin{tabular}{cc}
\includegraphics[width=0.45\textwidth]{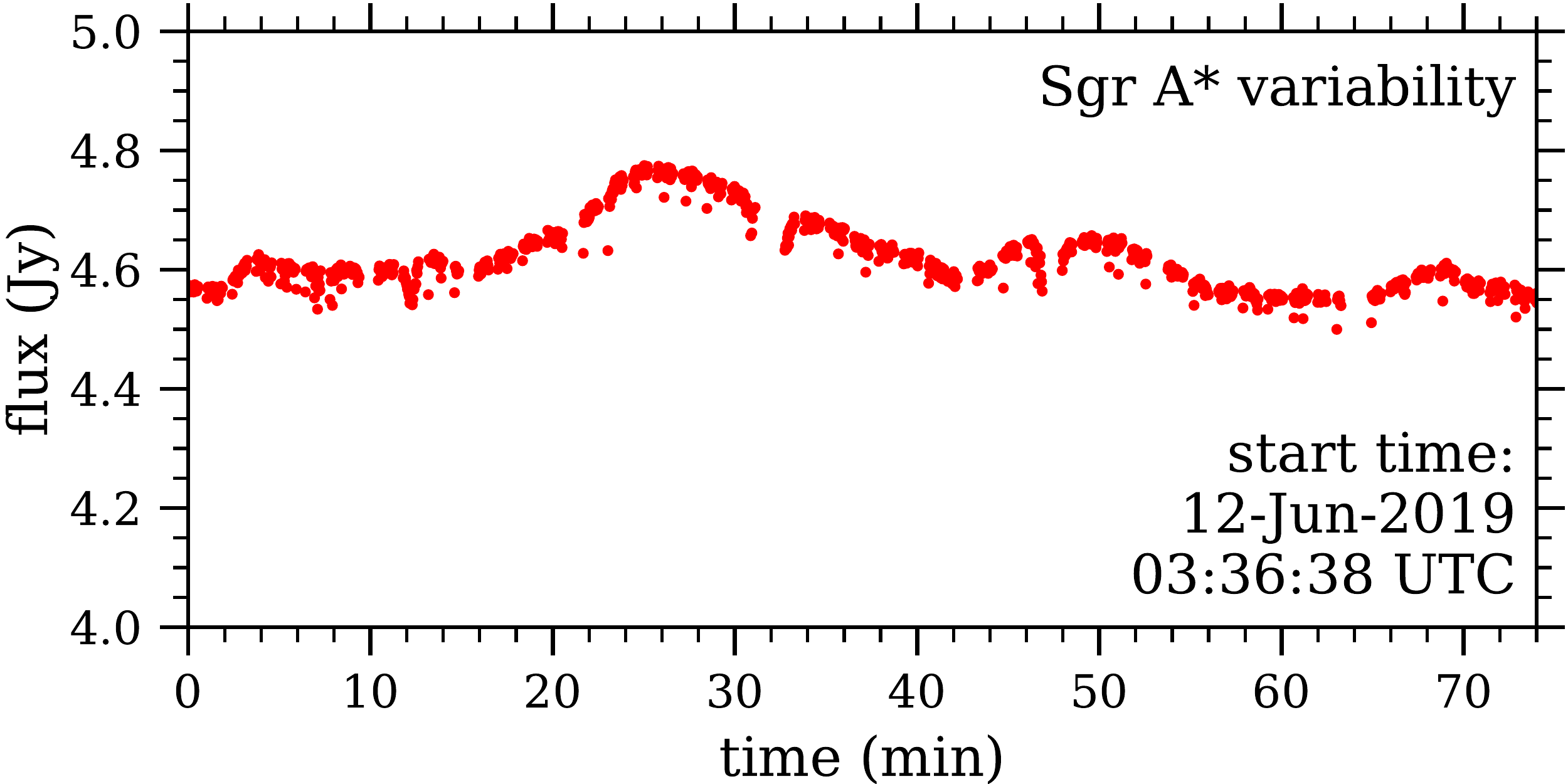} &
\includegraphics[width=0.45\textwidth]{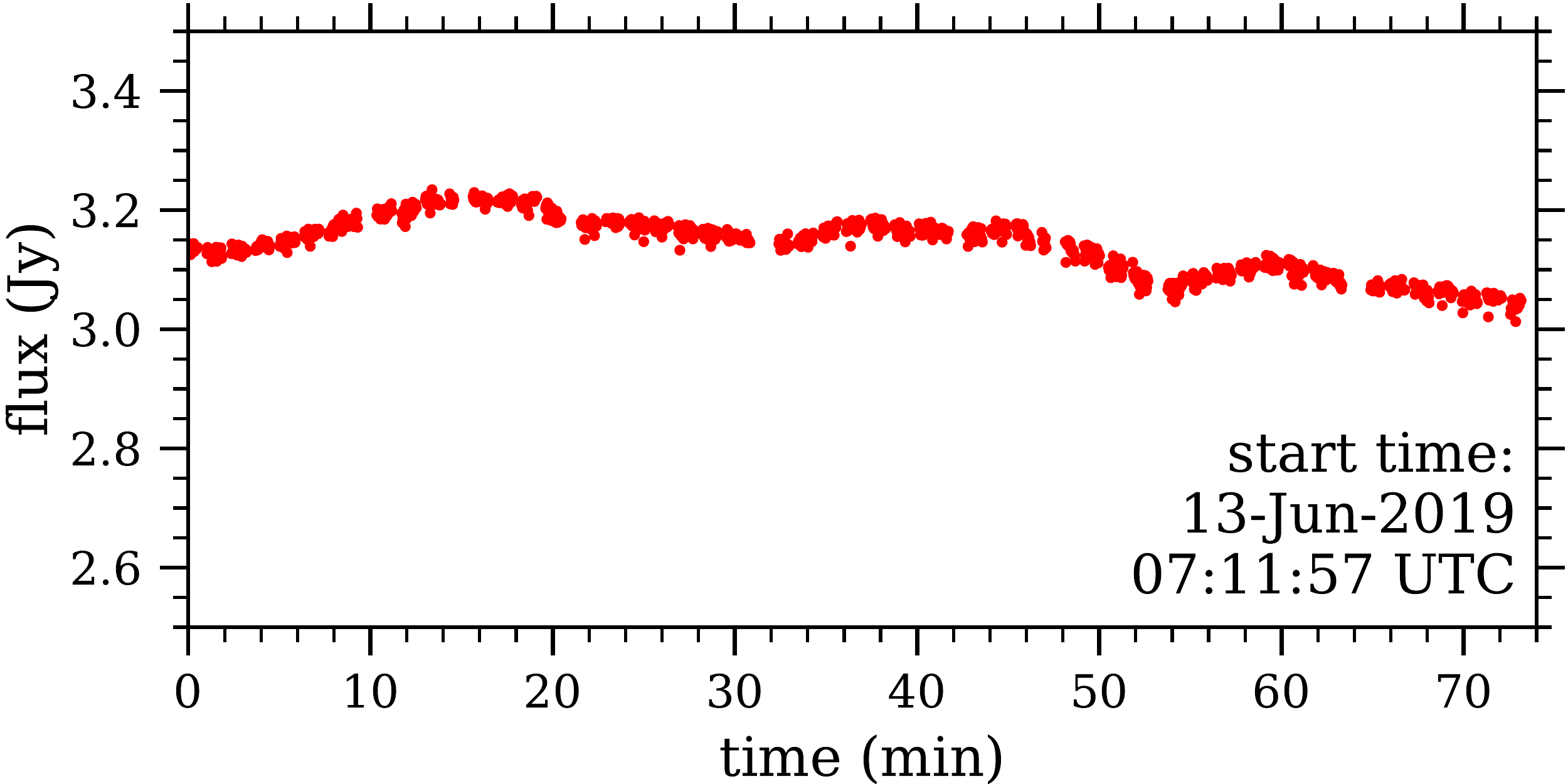}
\\
\includegraphics[width=0.45\textwidth]{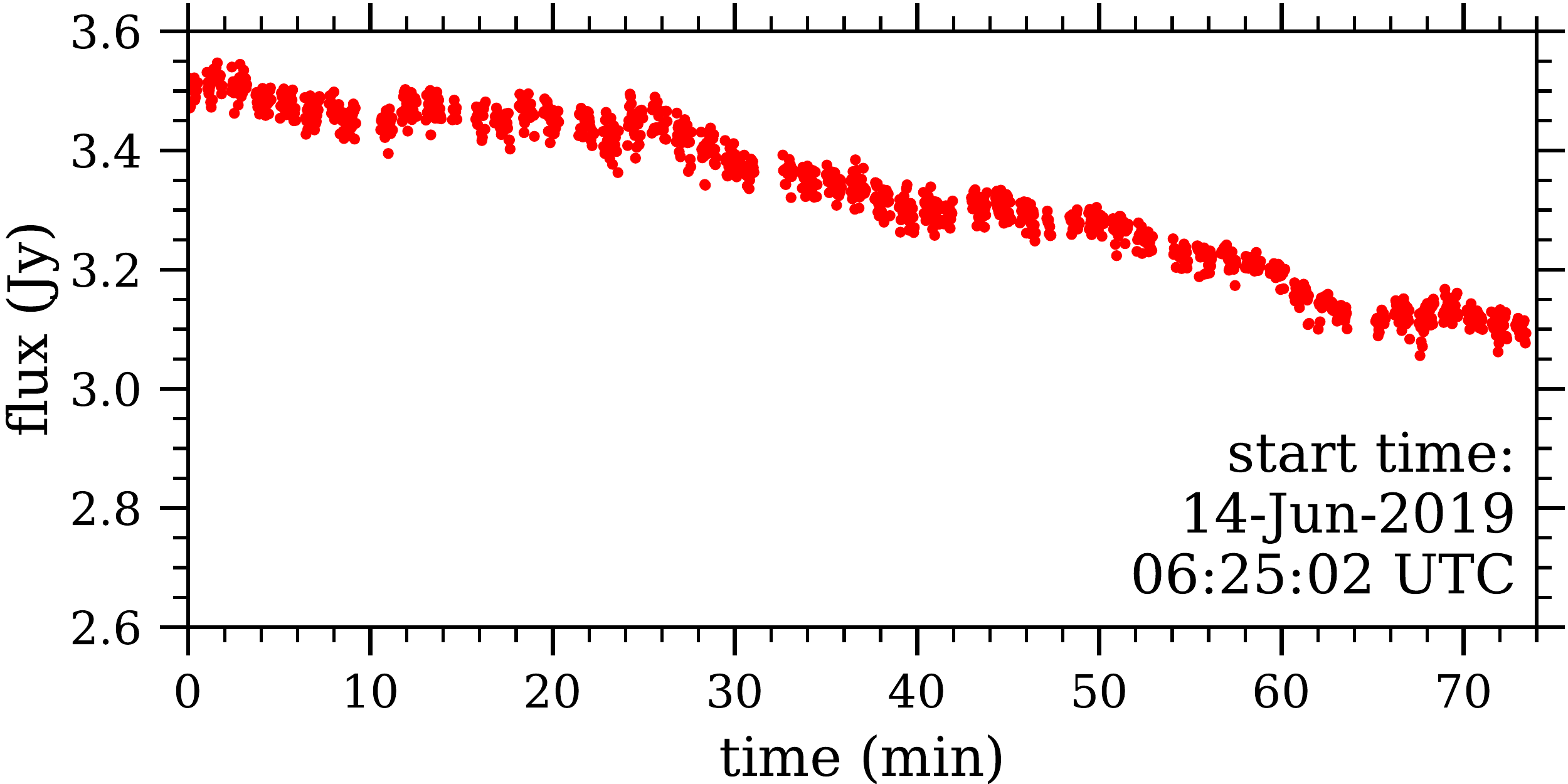} &
\includegraphics[width=0.45\textwidth]{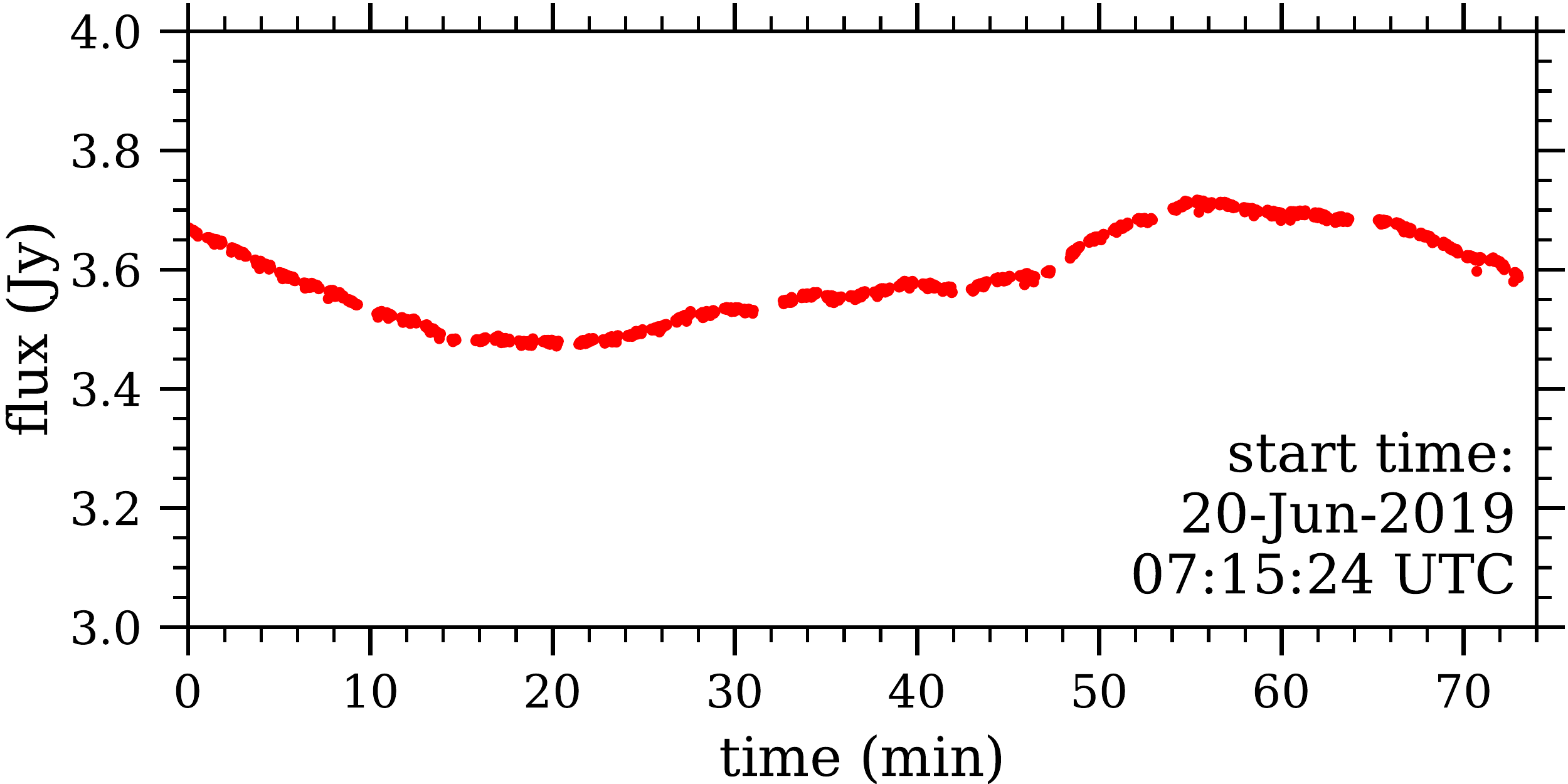}
\\
\multicolumn{2}{c}{\includegraphics[width=0.93\textwidth]{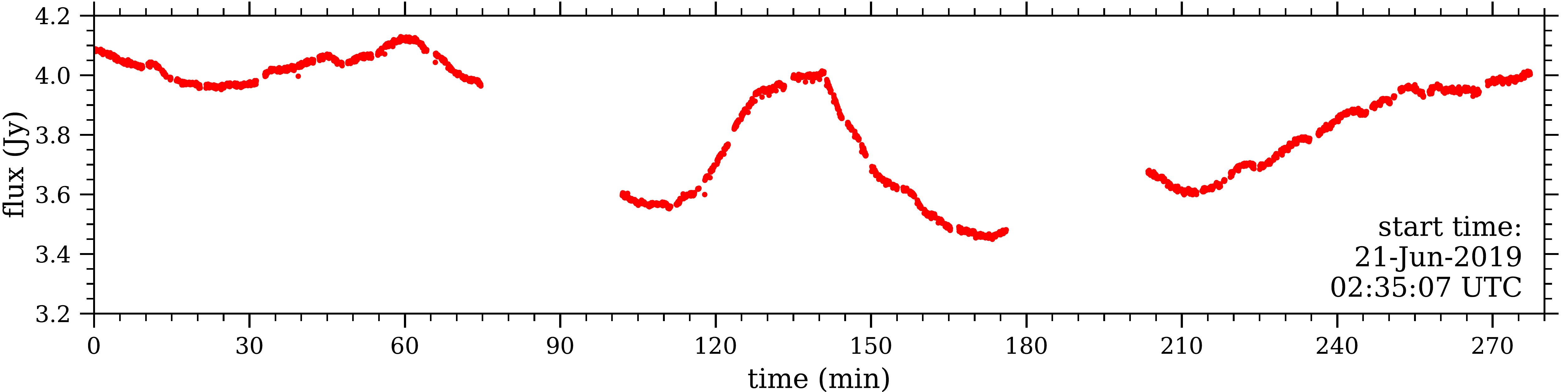}}
\end{tabular}
\caption {Variation of Sgr A* continuum flux at 230 GHz with time, i.e., light curves for our seven observations in ALMA Cycle 6. The observations are taken in June 2019, one month after the extraordinary flare of \cite{2019ApJ...882L..27D} and in the midst of the flaring activity of 2019. We averaged the data between the frequencies of 228.55~GHz and 233.8 GHz. The exact central frequency is 231.2 GHz. The start time of the observation is marked in the bottom right corner of each panel. Each point plotted represents a 2 second integration. Gaps in the light curves are due to observations of the calibrators. The three consecutive observations are plotted together on the bottom panel. Due to weather conditions, observational sensitivity varies from epoch to epoch. It is not uncommon to see occasional spurious features like in 12-Jun-2019 dataset, mostly at the ends of scans, which are sometimes related to a short term issues with a subset of antennas.}
\label{fig:lightcurves}
\end{figure*}

We develop a new algorithm for extracting variability information.
After phase self-calibration we collapse the data cube along the frequency axis using all of the line-free channels. We then split the $(u,v)$-data time stamp by time stamp and find the best fit point source model.\footnote{Fitting either a perfect point source or a Gaussian source with a variable width and position yields essentially identical result. In both cases the CASA task uvmodelfit successfully find the location of Sgr A*. This location does not drift within observations.}
This allows us to extract the light curves at the telescope's data output cadence, which is 2 seconds for these observations. The excellent ALMA sensitivity allows us to achieve a signal-to-noise ratio of about 500 per each data point. We extract 1359 data points for each 74 min observation (Figure \ref{fig:lightcurves}).
As $(u,v)$-model fitting occasionally fails on the scale of a few percent, we remove points which are outstanding more than 1-1.5\%
%1\% (or 1.5\% for the June 14, 2019 dataset) 
from the average of its ten closest neighbors.
%54 points per each observation are discarded because the last data point in every scan is always an outlier. 
The total number of removed points is between 54 and 140 per observation, i.e. 4-10\%.

The emission from the Galactic Center minispiral is largely resolved out in our observations. Its total integrated emission across the whole field of view is about 0.25 Jy. This extended emission has different $(u,v)$-signature compare to the central point source. It contribute to the uncertainty of the model fit on the scale of  ~0.4 mJy, which is smaller than our observational uncertainties. In cases of low resolution data and/or to improve on the precision of the light curves it is important to subtract the $(u,v)$-signature of the minispiral before extracting light curves.

Typically, variability studies using ALMA data employs imaging. Reliable imaging requires binning data into about 1-1.5 minute blocks. Consequently, the time resolution of Sgr A* light curves obtained with ALMA in the literature are about 30-50 times lower than ours. The highest cadence light curves published in the literature are from the Submillimeter Array (SMA), and have a time resolution of 15 seconds. However, on average, SMA's signal-to-noise ratio is considerably lower than ALMA's. The latter is also the case for 
%CARMA, SMA, IRAM and 
other telescopes referenced here.

Our algorithm utilizes the highest cadence permitted by the telescope and all available $(u,v)$-data. To avoid the light curve contamination with extended emission we use $(u,v)$-model fitting. 
We therefore avoid unnecessary imaging of data on short time intervals which tends to be effected by artifacts and results in higher uncertainties for light curve points. The downside of our algorithm is that it is computationally expensive.

\section{Results and Intrinsic structure function}\label{sec:results}

\begin{figure}
\vspace{-0.0cm}
\centering
\includegraphics[width=0.45\textwidth]{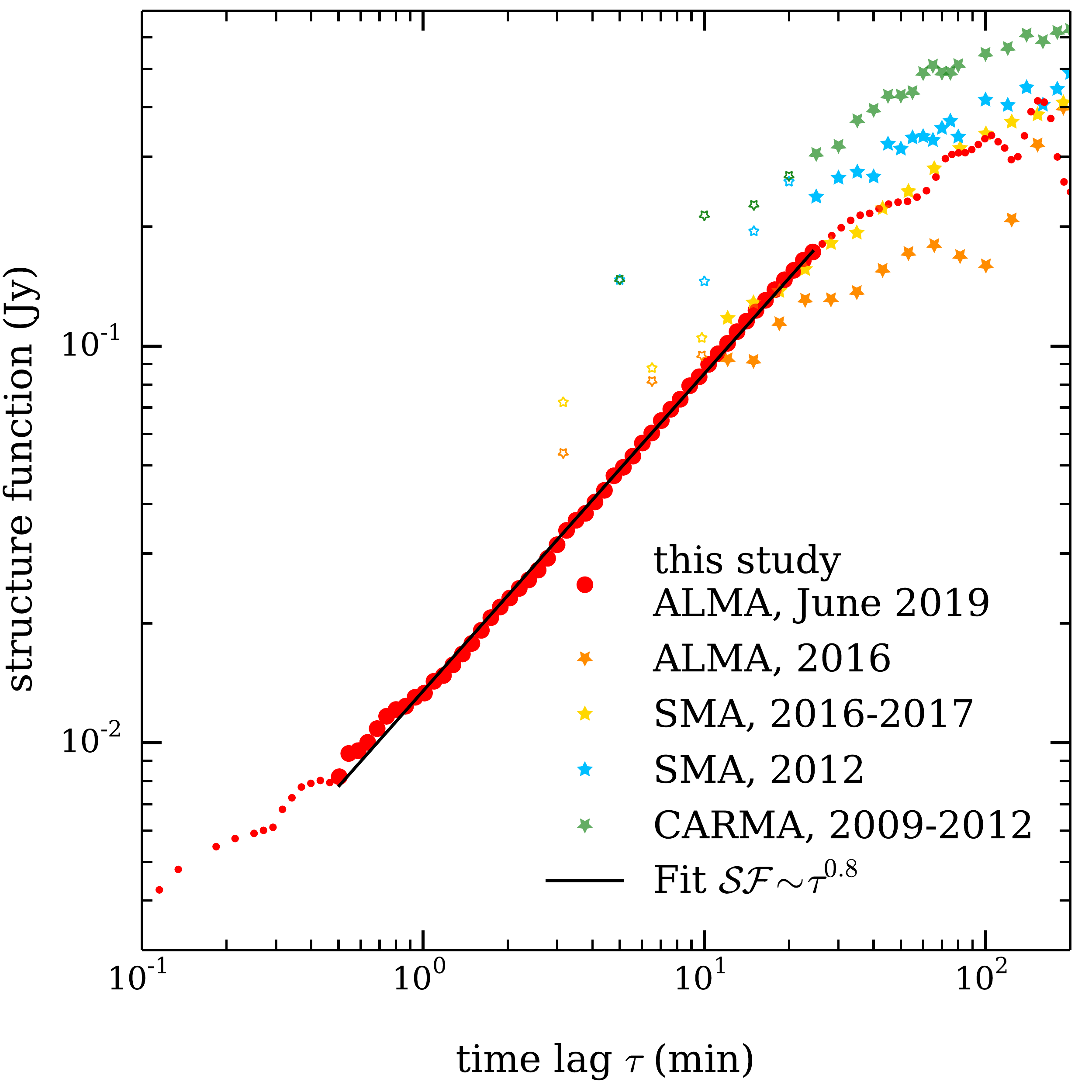}\\
\caption {Comparison of intrinsic structure functions of Sgr A* variability at 230 GHz. The June 2019 $\mathcal{SF}(\tau)$ in in red. Large circles represent statistically reliable time range $30\mathrm{sec}<\tau<25\mathrm{min}.$ The black line is the power-law fit. Small circles represent low statistics points at $\tau > 25$ min and $\tau < 30$ sec where the noise subtraction was less reliable.
The intrinsic structure functions obtained with data from ALMA, SMA and CARMA at various epochs is presented with stars. Empty stars represent the points where the observational uncertainties could not be reliably subtracted.  ALMA data from 2016 (orange) and SMA data from 2016-2017 (yellow) are from  \cite{Witzel2020}. 
We removed their only 2015 dataset due to its high level of noise.
SMA data from 2012 (blue) and CARMA data from 2009-2012 (green) are from \cite{Dexter2014}.
The plotted intrinsic structure functions are obtained by subtracting known uncertainties. The remaining uncertainties are primarily statistical, i.e. due to the finite size of the dataset and an accuracy of determining the observational uncertainties. The latter is dominating the uncertainties on our combined  $\mathcal{SF}(\tau)$, we estimate $\sigma_{\mathcal{SF}} \simeq 0.1 \langle \sigma_{obs} \rangle \sim 10^{-3}$ Jy, where $\langle \sigma_{obs} \rangle$ is the mean observational uncertainty.
}
\label{fig:structure.function}
\end{figure}

\begin{figure}
\vspace{-0.0cm}
\centering
\includegraphics[width=0.45\textwidth]{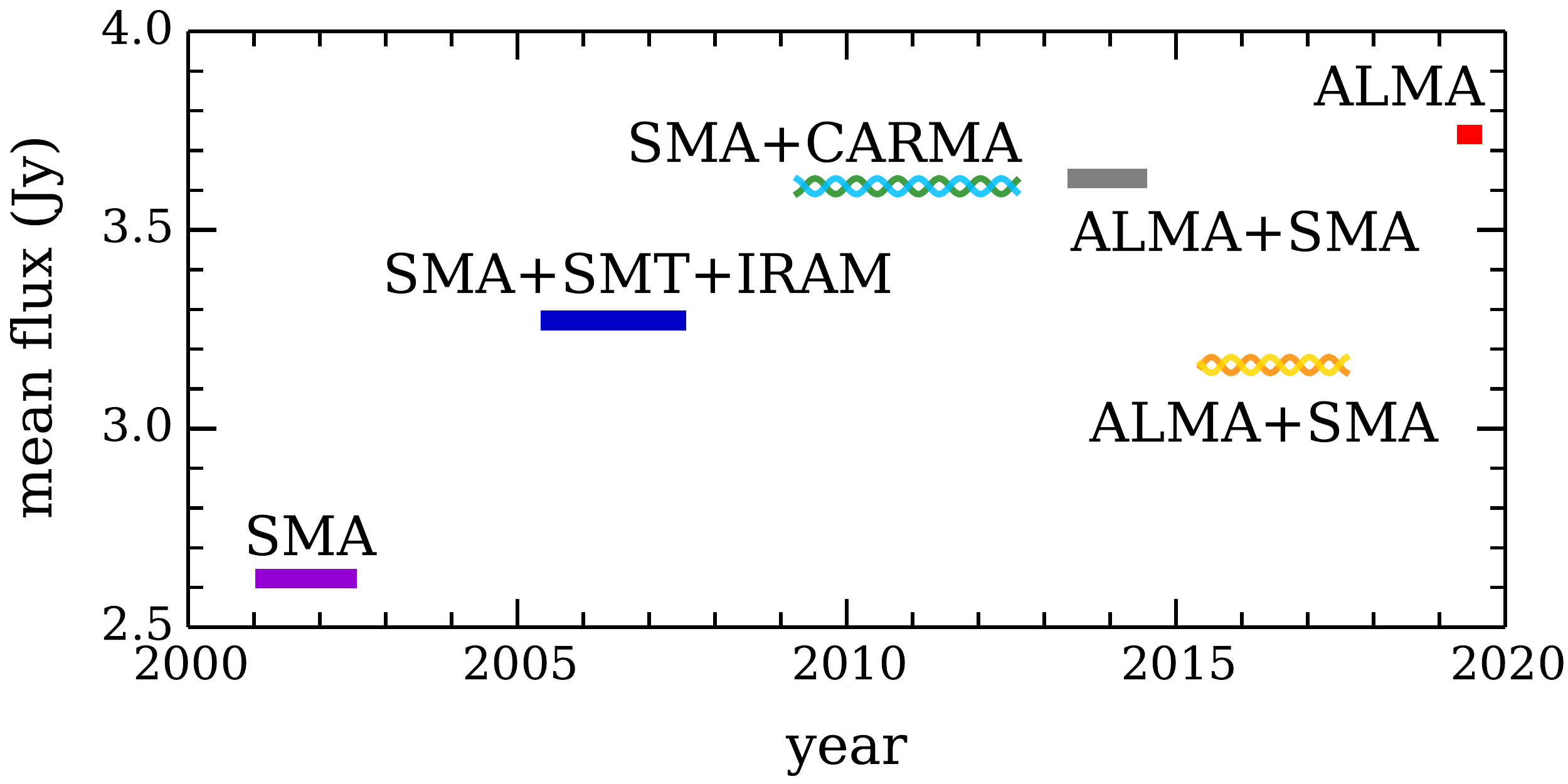}
\caption {Comparison of mean flux densities of Sgr A* at 230 GHz. Red: June 2019 data (red). Intertwined orange and yellow: 2015-2017 data from \cite{Witzel2020}. Grey: 2013-2014 data from \cite{Bower2015}. Intertwined blue and green: 2019-2012 data from \cite{Dexter2014}. Dark blue: Combined 2005-2007 data from \cite{2008ApJ...682..373M} and \cite{2009ApJ...706..348Y}. 
Violet: 2001-2002 data from \cite{2003ApJ...586L..29Z}. The telescopes which obtained the data are listed next to the mean flux density. We combine the datasets such that there are about 10 epochs in each. Typical standard deviation is $\sim 0.5$~Jy.  To calculate the mean of data from \citet{Bower2015} we average their eight ALMA data points and the 16 SMA data points at 226.85 GHz.
}
\label{fig:means}
\end{figure}

Figure \ref{fig:lightcurves} shows the 230 GHz variability of Sgr A* in June 2019. Average flux across our observations is 3.74 Jy. % with the standard deviation 0.46 Jy. 
The minimum and maximum across all observations, calculated as an average of ten neighboring points, are 3.04 Jy and 4.77 Jy.

To quantify variability properties we construct the structure function (or a square root of variance) 
\be
	\mathrm{SF}(\tau)=\sqrt{\frac{1}{N_\tau}\sum\limits_{\mathrm{pairs}}(F(t+\tau)-F(t))^2},
	\label{eq:sf}
\ee
where $\tau$ is the time lag, $F(t)$ is the flux at time t, $N_\tau$ is the number of pairs of points separated by the time interval $\t$ in the data, and the summation runs over all such pairs. 

At small timescales the structure function computed this way tends to be dominated by observational uncertainties. It is easily seen from flattening of the structure functions found in the literature (e.g. \citealt{Witzel2020,Iwata2020,Dexter2014}) at the level an average observational uncertainty. 

Let us introduce the \textit{intrinsic structure function} $\mathcal{SF}$, which can be computed from observed structure function ($\mathrm{SF}$) by removing the contribution of observational uncertainties
\be
    \mathcal{SF}(\tau)=
    \sqrt{\frac{1}{\sum\limits_{obs} N^{obs}_\tau}
    \sum\limits_{obs} \left[
    \mathrm{SF}_{obs}^2(\tau)-2 \s_{obs}^2
    \right]^2 
    N_\tau^{obs}}.
    \label{eq:sf.tot}
\ee
Here $\mathrm{SF}_{obs}$ is a structure function of each individual observation, $N^{obs}_\tau$ is a pair count within each individual observation, and the summation runs over independent observations.
We treat the consecutive observations 4, 5, and 6 as one observation. 

To derive equation \ref{eq:sf.tot} we notice that in the presence of observational noise the observed flux $F(t)$ at time $t$ can be decomposed into the sum of the true flux $\mathcal{F}(t)$ and a contribution of observational noise $\d f(t)$ such that $F(t)=\mathcal{F}(t)+\d f(t).$ Then the structure function obtained using the data from each observation $\mathrm{SF}_{obs}(\t)$ can be re-written as (see also \citealt{1985ApJ...296...46S})
\be
	\mathrm{SF}_{obs}^2(\tau)
	&& ={\frac{1}{N_\tau}\sum\limits_{\mathrm{pairs}}(\mathcal{F}(t+\tau)+\d f (t+\tau)-\mathcal{F}(t)- \d f (t))^2}\\
	&& = {\frac{1}{N_\tau}\sum\limits_{\mathrm{pairs}} (\mathcal{F}(t+\tau)-\mathcal{F}(t))^2
	+ \frac{1}{N_\tau}\sum\limits_{\mathrm{pairs}}\d f^2(t+\tau)
	+ \frac{1}{N_\tau}\sum\limits_{\mathrm{pairs}} \d f^2(t)}\\
	&& ={\mathcal{SF}_{obs}^2(\tau)
	+ 2 \sigma_{obs}^2} \, , \label{eq:sf.dir}
\ee
where $\mathcal{SF}_{obs}^2(\tau) = {\frac{1}{N_\tau}\sum\limits_{\mathrm{pairs}}(\mathcal{F}(t+\tau)-\mathcal{F}(t))^2}$ is the intrinsic structure function and
$\sigma_{obs}^2={\frac{1}{N_\tau}\sum\limits_{\mathrm{pairs}}\d f^2(t)}
$ is the mean-square root observational uncertainty. The equation \ref{eq:sf.dir} can be recast as equation \ref{eq:sf.tot} for multiple observations.

The derivation holds precisely only for a large sample size, as it relies on the fact that an expectation value of two independent distributions $\mathcal{F}$ and $\delta f$ is the product of expectation values for these distributions individually: $\langle \mathcal{F} \delta f \rangle= \langle \mathcal{F} \rangle \langle \delta f \rangle = 0,$ if $\langle \delta f \rangle =0.$ %Here $\delta F$ stands for the observational uncertainty. 
In case of a small sample size, as is the case of all variability studies so far, the observational uncertainties can not be subtracted completely. 

The resulting structure function is plotted in Figure \ref{fig:structure.function}.
We find that the structure function for $\tau < 25$ min scales approximately as $\tau^{0.8}$.
At $\tau \sim 2$ seconds the imperfection of determining $\sigma_{obs}$ and lack of statistics (which is still the case even in this comparatively large sample) influences the result. At longer timescales $\tau \geq 25$ min the size of our sample begins to influence the results, as the number of independent $(F(t+\tau), F(t))$ pairs decreases, and the sample is dominated by the three consecutive June 21, 2019 observations. We still present the structure function at larger timescales in Figure \ref{fig:structure.function}.

\section{Discussion and Comparison}\label{sec:discussion}

We compare our results with previous studies of Sgr A* variability at 230 GHz. \cite{Witzel2020} use the 3000 min of data observed in 2015-2017 and include the data sample of  \cite{Iwata2020}. \cite{Bower2015} use the data observed in 2013-2014. \cite{Dexter2014} use the data observed in 2001-2012. The latter study is over the largest time span, however the data are sparse and irregularly sampled, and consists of a total of 1044 data points.
%at 230 GHz distributed over 29 epochs with the shortest cadence of 5 and 10 min. 
%while, in this study, we present more than 8400 points.

Since Sgr A* is a highly variable source and in the view of the sparsity of observations there is little value in comparing the minimum and maximum fluxes encountered during the observations. These values are highly volatile and they are additionally affected by the absolute flux level calibrations\footnote{Correcting for the uncertainty in the absolute flux-density calibration is very important in computing the Sgr A* structure function on long time scales, particularly those involving approximations between data sets.}. Instead, we compare global statistical measures such as mean values and structure function values. 

The comparison of mean flux densities over the last 20 years is presented in Figure \ref{fig:means}.  In June 2019 we obtained a mean flux value of 3.74 Jy. This is 20\% higher than in the epoch directly preceding it \citep{Witzel2020}. 
It is however only 3\% higher than in 2009-2012 and in 2013-2014 \citep{Bower2015,Dexter2014}. The submm flux is dominated by a population of thermal electrons and is expected to vary together with the mean accretion rate \citep{Yuan2014,Bower2019}. There is a slow variability of the mean Sgr A* flux in the submm on the scale of $\sim$10 years (Figure \ref{fig:means}) which is similar to expected global mass accretion variability \citep{2020MNRAS.492.3272R}. Therefore we deduce that the mean accretion rate in June 2019 is about 20\% higher than in 2015-2017, and about the same as in 2009-2012.

The comparison of intrinsic structure functions of Sgr A* $\mathcal{SF}$ defined in equation \ref{eq:sf.tot} with the one derived in this work is shown in Figure \ref{fig:structure.function}. 
We use \cite{Witzel2020} and \cite{Dexter2014} (included with the publications) data to recalculate the corresponding intrinsic structure functions at earlier epochs. Due to low statistics and the fact that at the shortest timescales the $\mathcal{SF} \sim \s_{obs},$ the earlier intrinsic structure functions are still affected by observational uncertainties at timescales $<10$ min. 
%We remove the noisiest 2015 SMA data set because its high noise level overwhelms the structure function at timescales below a tenth of minutes. 
No variability comparison is possible with \cite{Bower2015}'s 2013-2014 data, as they have only one epoch of monitoring.

The slope of the 2019 structure function of 0.8 is consistent with red noise \citep{2010MNRAS.404..931E} and 1-$\sigma$ consistent with the power spectrum determined by \cite{Witzel2020}. The high signal-to-noise ratio of the high cadence observations of 2019 allows us to demonstrate for the first time that the red noise characteristics of the mm-variability of Sgr~A* continue down to a time lag of 30~seconds. The red noise behavior seems to be intrinsic to the source. The overall level of Sgr A* variability in June 2019 seems to be the same or somewhat higher than that of 2016-2017 and somewhat lower than that of 2009-2012. An accurate comparison of the variabilities is complicated by the systematic differences among the derived historic structure functions, some of which may be due to higher uncertainties in observations in 2000s.

\section{Conclusion}\label{sec:conclusion}

We present ALMA observations of Sgr A* variability at 230 GHz in June 2019 (Figure \ref{fig:lightcurves}). ALMA's excellent sensitivity together with a new light curve extraction algorithm developed for this work allows us to achieve a combination of high cadence (2 sec) and high signal-to-noise ratio ($\sim 500$), which surpasses light curves available in the literature by a factor of 10-100.

We construct an intrinsic structure function of Sgr A* variability at 230 GHz by subtracting the contribution of observational uncertainties as described in Section \ref{sec:results}, with the result shown in Figure \ref{fig:structure.function}.
We push the range of validity of the structure function by about two orders of magnitude toward shorter timescales.
%The reliability range of the structure function obtained in this work surpasses previous studies by a factor of about hundred in shortness of timescale and a factor of ten in sensitivity.
We find that it can be approximated by 
the power-law $\mathcal{SF}(\tau)\sim \tau^{0.8}$ on timescales $30 \, \mathrm{sec}\leq \tau \leq 25 \, \mathrm{min}$.

Our observations are taken one month after the brightest NIR flare of \cite{2019ApJ...882L..27D} and in the midst of the NIR flaring activity of 2019. Using the connection between the submm flux of Sgr A* and the accretion rate \citep{Yuan2014,Bower2019}, we conclude that the flaring activity of 2019 coincides with the period of elevated accretion rate onto the black hole as compared to the epoch directly preceding it \citep{Witzel2020}, in agreement with the calculation of \cite{Murchikova2021}. 

We find that the elevated accretion rate was not the trigger of the bright NIR flares observed in 2019 \citep{2019ApJ...882L..27D,Gravity2020}. An almost identical mean submm continuum flux and consequently the identical mean accretion rate was observed between 2009 and 2014 \citep{Dexter2014,Bower2015} when no bright flaring activity was reported (Figure \ref{fig:means}). Bright NIR flares in general do not get brighter with increased accretion rate, hence their relation
to the accretion rate, if any, must be indirect.

Among the physical mechanisms proposed as possible origins of the NIR flares such as population of non-thermal electrons, magnetic reconnection and shocks \citep{Dodds-Eden2010,2017MNRAS.468.2447P,2012AJ....144....1Y,2003ApJ...598..301Y},
the one the most independent from accretion rate is magnetic reconnection
\citep{Ripperda2020,Ripperda2021}. We therefore suggest that the brightest NIR flares of Sgr A* are likely caused by magnetic reconnection.
We stress that this statement is related only to the brightest among the NIR flares of Sgr A*. The full phenomenology of the variability is likely produced by a combination of the mechanisms listed above.

\acknowledgments
%\section*{Acknowledgments}

We are grateful to Jason Dexter, Mark Gurwell, Brian Mason, Sasha Philippov, Eduardo Ros, Nadia Zakamska, and the anonymous referee for comments and suggestions.
LM's membership at the Institute for Advanced Study is supported by the Corning Glass Works Foundation. A part of this work was performed at the Aspen Center for Physics, which is supported by National Science Foundation grant PHY-1607611. The participation of LM at the Aspen Center for Physics was supported by the Simons Foundation.

This paper makes use of the following ALMA data:
  ADS/JAO.ALMA\#2018.1.01124.S. ALMA is a partnership of ESO (representing
  its member states), NSF (USA) and NINS (Japan), together with NRC
  (Canada) and NSC and ASIAA (Taiwan) and KASI (Republic of Korea), in
  cooperation with the Republic of Chile. The Joint ALMA Observatory is
  operated by ESO, AUI/NRAO and NAOJ.

The National Radio Astronomy Observatory is a facility of the National Science Foundation operated under cooperative agreement by Associated Universities, Inc.

%A handy "cheat sheet" that provides the necessary LaTeX to produce 17 
%different types of tables is available at \url{http://journals.aas.org/authors/aastex/aasguide.html#table_cheat_sheet}.




%\begin{thebibliography}{}

\bibliography{sgra_flares}{}
\bibliographystyle{aasjournal}

%\end{thebibliography}



\end{document}